\documentclass{article}
\usepackage{spconf,amsmath,graphicx,hyperref}
 \usepackage{booktabs}
 \usepackage{xcolor}

\title{AudioFuse: Unified Spectral-Temporal Learning via a Hybrid ViT-1D CNN Architecture for Phonocardiogram Classification}
\name{Md. Saiful Bari Siddiqui, Utsab Saha\thanks{Thanks to BRAC University for institutional support.}}
\address{Department of Computer Science and Engineering, BRAC University, Dhaka, Bangladesh}
\begin{document}
%
\maketitle
\begin{abstract}
Biomedical audio signals, such as phonocardiograms (PCG), are inherently rhythmic and contain diagnostic information in both their spectral (tonal) and temporal domains. Standard 2D spectrograms provide rich spectral features but compromise the phase information and temporal precision of the 1D waveform. We propose AudioFuse, an architecture that simultaneously learns from both complementary representations to classify PCGs. To mitigate the overfitting risk common in fusion models, we integrate a custom, wide-and-shallow Vision Transformer (ViT) for spectrograms with a shallow 1D CNN for raw waveforms. On the PhysioNet 2016 dataset, AudioFuse achieves a state-of-the-art competitive ROC-AUC of 0.8608 when trained from scratch, outperforming its spectrogram (0.8066) and waveform (0.8223) baselines. Moreover, it demonstrates superior robustness to domain shift on the challenging PASCAL dataset, maintaining an ROC-AUC of 0.7181 while the spectrogram baseline collapses (0.4873). Fusing complementary representations thus provides a strong inductive bias, enabling the creation of efficient, generalizable classifiers without requiring large-scale pre-training.
\end{abstract}
\begin{keywords}
Acoustic signal classification, Biomedical signal processing, Convolutional neural networks, Representation learning, Vision transformers.
\end{keywords}
\section{Introduction}
\label{sec:intro}

The analysis of biomedical audio signals is a critical and rapidly growing field, offering the potential for low-cost, non-invasive, and scalable diagnostic tools. A significant portion of these signals, such as phonocardiograms (PCG), are inherently rhythmic. Their diagnostic value is encoded not just in their spectral content (such as the frequency of a heart murmur), but also in their fine-grained temporal dynamics, including the precise timing, duration, and morphology of acoustic events like S1/S2 heart sounds. The dominant paradigm for biomedical audio classification involves converting the 1D time-series signal into a 2D representation\cite{gong2021ast}, typically a log-Mel spectrogram, and applying powerful vision-based deep learning models like CNNs or Transformers~\cite{liu2024efficient, banerjee2020semi, zhao2025detection, sharma2025spectronet}. While this approach excels at learning complex spectral and harmonic patterns, it creates an inherent trade-off. The windowing function of the Short-Time Fourier Transform (STFT) used to generate spectrograms necessarily blurs temporal details and discards all phase information, potentially obscuring the very transient and rhythmic cues that are diagnostically critical~\cite{BinhThien2023},~\cite{GriffinSTFT1984}. Conversely, models operating directly on the 1D waveform preserve this temporal precision but struggle to learn the complex frequency relationships that are explicit in a 2D representation~\cite{NatsiouAudio2021},~\cite{PeetersRichard2021}.

To address this limitation, we propose \texttt{AudioFuse}, a dual-branch architecture designed to simultaneously learn from these complementary spectral and temporal representations. A key challenge with fusion models is the heightened risk of overfitting, due to a potentially large parameter count and increased complexity, especially on the limited-sized biomedical domain datasets~\cite{Guarrasi2025}. To mitigate this, we specifically design \texttt{AudioFuse} to be lightweight and parameter-efficient. It employs a balanced two-branch design: a custom, wide-and-shallow ViT processes spectrograms to learn global spectral context, while a compact 1D-CNN extracts precise temporal features from the raw waveform. These specialized feature vectors are then integrated via late fusion.

We validate our approach on the challenging task of PCG classification, using the PhysioNet 2016 dataset to evaluate in-domain performance and the PASCAL dataset to test for out-of-domain generalization. Our experimental results show that \texttt{AudioFuse}, when trained from scratch, significantly outperforms its single-modality baselines, achieving SOTA-competitive results despite being lightweight. It also demonstrates substantially improved generalization performance. 
 
\section{Theoretical Background}
\label{sec:background}

\subsection{Complementary 1D and 2D Audio Representations}
\label{ssec:complementarity}

Let a discrete audio signal be represented as a 1D time-series vector $x[n]$, where $n$ is the sample index. This representation preserves all information, including phase and precise temporal location of events.

Conversely, a 2D log-Mel spectrogram, $S(t, f_{mel})$, is a derived representation. It is generated via the Short-Time Fourier Transform (STFT), which maps the signal from the time domain to a joint time-frequency domain:
\begin{equation}
\label{eq:stft}
X(t, \omega) = \sum_{n=-\infty}^{\infty} x[n] w[n-t] e^{-j\omega n}
\end{equation}

where $w[n]$ is a windowing function. 
The power spectrogram is then $P(t, \omega) = |X(t, \omega)|^2$, which is subsequently mapped to the Mel scale. The fundamental trade-off here is that to achieve high frequency resolution, the window $w[n]$ must be long. This, however, averages the signal over a longer time period, resulting in poor time resolution and blurring the precise onset of transient events like cardiac valve clicks. Conversely, a short window provides good time resolution but poor frequency resolution. This highlights the complementary nature of the two representations:
\begin{itemize}
    \item \textbf{1D Waveform ($x[n]$):} Possesses high temporal resolution and complete phase information. It is ideal for localizing transient events and analyzing rhythm. Its spectral information, however, is only implicit.
    \item \textbf{2D Spectrogram ($S(t, f_{mel})$):} Possesses high spectral resolution, making harmonic structures visually explicit. It is ideal for analyzing tonal components like murmurs or wheezes. Its temporal information is inherently compromised.
\end{itemize}
No single representation is optimal for all diagnostic features present in a biomedical signal. Therefore, a model that can leverage the strengths of both is theoretically superior. Several approaches ranging from traditional ML~\cite{Zeinali2021} to Encoder-Decoder architectures~\cite{banerjee2020semi} have been explored for heart sound classification by researchers, but very few works have been done to combine these two complementary domains.

\subsection{Feature Extraction and Fusion}
\label{ssec:architectures}

To effectively learn from these two disparate representations, we require specialized network architectures for each branch. For the image-like spectrogram, the \textbf{ViT}\cite{dosovitskiy2020image} is exceptionally well-suited. Its self-attention mechanism\cite{vaswani2017attention} allows it to learn global, long-range dependencies across the entire time-frequency plane, enabling it to model the overall structure and evolution of harmonic patterns more effectively than the local receptive fields of a traditional CNN.

For the sequential waveform, a \textbf{1D-CNN} is an ideal choice. Its 1D kernels act as learnable matched filters that slide along the time axis, making them highly efficient at detecting specific, short-duration temporal patterns and wave morphologies that characterize transient events. Given that the two representations contain fundamentally different and complementary types of information, it is crucial to allow each specialized network to extract a high-level feature representation independently, without premature mixing. Therefore, we adopt a \textbf{late-fusion} strategy~\cite{Ngiam2011}. The ViT and 1D-CNN branches act as independent feature extractors, each processing its preferred modality. The final, high-level feature vectors from each branch are then fused. This allows a subsequent classification head to learn the final, complex decision boundary based on the summarized, high-level insights from both the spectral and temporal domains, preserving the integrity of each feature extraction process.

\section{Proposed Method: AudioFuse}
\label{sec:method}

\begin{figure*}[ht]
  \centering
  \includegraphics[width=0.85\linewidth]{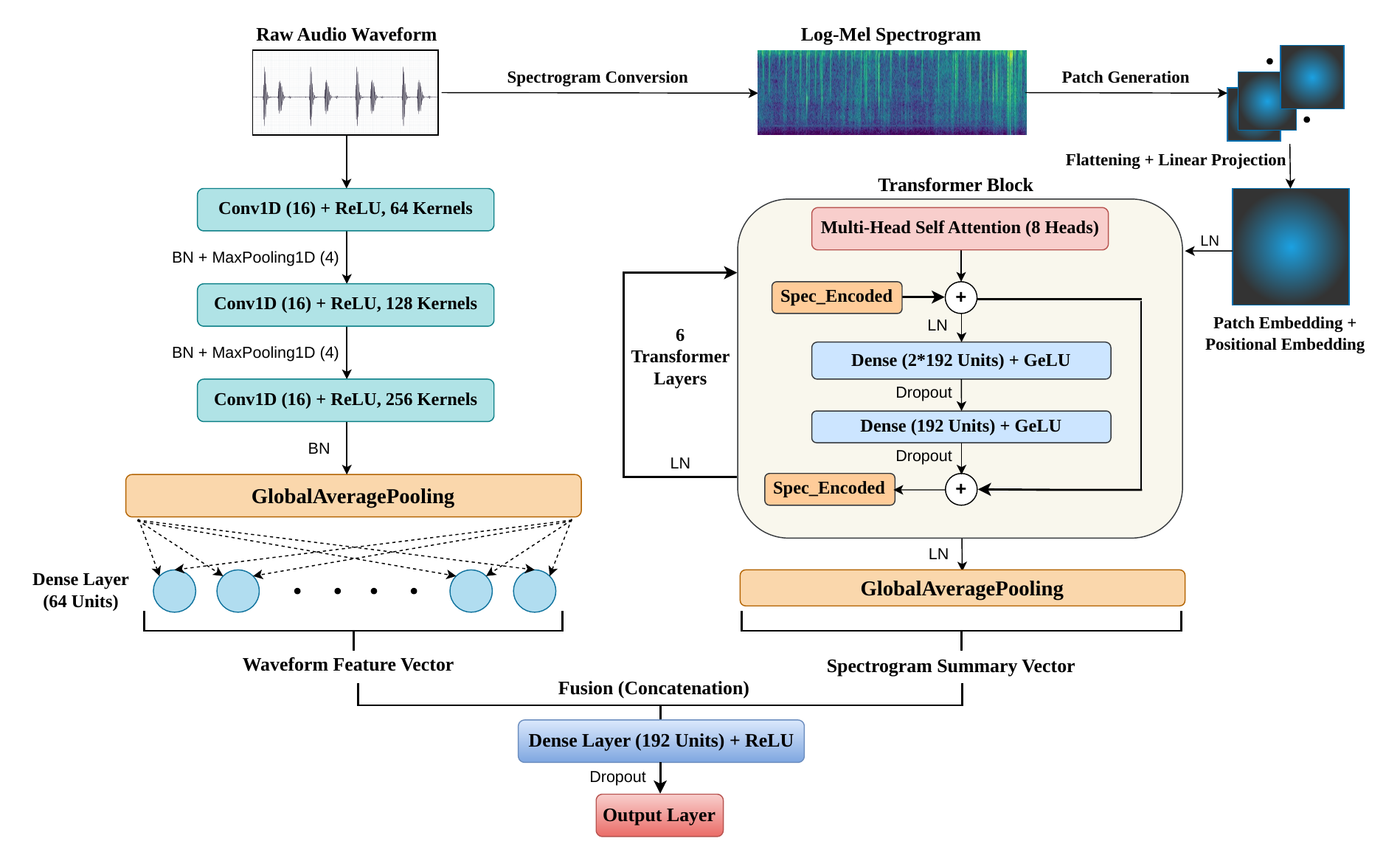} 
  \caption{The AudioFuse architecture. The 2D spectrogram and 1D waveform are processed by parallel ViT and 1D-CNN branches, respectively. The resulting feature vectors are concatenated and passed to a final MLP for classification.}
  \label{fig:architecture}
\end{figure*}

Our proposed \texttt{AudioFuse}, a lightweight, dual-domain fusion architecture, is shown in Fig.~\ref{fig:architecture}.

\subsection{Spectrogram Branch}
\label{ssec:vit_branch}

The spectral branch is designed to learn global context from the harmonic and tonal structures within a 2D log-Mel spectrogram. For this task, we designed a custom wide-and-shallow Vision Transformer (ViT). The input $224 \times 224$ spectrogram is first transformed into a sequence of 196 patch embeddings, each with a dimensionality of 192, via a \texttt{Conv2D} layer with a $16 \times 16$ kernel. After learnable positional embeddings are added, this sequence is processed by a stack of 6 Transformer encoder blocks. Each block employs a Multi-Head Self-Attention layer with 8 heads and a feed-forward MLP. The final output sequence is then globally averaged to produce a single 192-dimensional spectral feature vector, $\mathbf{f}_{spec}$. This shallow design captures essential global patterns while minimizing the parameter depth to prevent overfitting.

\subsection{Waveform Branch}
\label{ssec:cnn_branch}

The temporal branch is designed to learn precise, timing-based features directly from the raw 1D audio waveform, which is padded or truncated to a fixed length of 5 seconds (110,250 samples). We employ a compact 1D Convolutional Neural Network (CNN) for this task. The network consists of three sequential \texttt{Conv1D} blocks with increasing filter sizes (64, 128, 256) and a kernel size of 16. Each block uses a stride of 4 and is followed by Batch Normalization and MaxPooling, allowing the model to learn hierarchical temporal patterns at multiple resolutions. After the final block, the feature sequence is globally averaged and passed through a final \texttt{Dense} layer with 64 units, to produce a compact 64-dimensional temporal feature vector, $\mathbf{f}_{wave}$.

\subsection{Late Fusion Head}
\label{ssec:fusion_head}

The two specialized feature vectors are integrated for the final classification employing late fusion. The spectral feature vector ($\mathbf{f}_{spec}$) and the temporal feature vector ($\mathbf{f}_{wave}$) are concatenated to form a single 256-dimensional fused vector:
\begin{equation}
\mathbf{f}_{fused} = [\mathbf{f}_{spec} ; \mathbf{f}_{wave}]
\end{equation}
This vector is then passed through an MLP head consisting of a single \texttt{Dense} layer with 192 units, a high-rate (0.5) \texttt{Dropout} layer for regularization, and a final sigmoid output layer, enabling the learning of a decision boundary based on the high-level, complementary insights from both branches.
\section{Results and Discussion}
\label{sec:results}

\subsection{Datasets and Implementation Details}
\label{ssec:implementation_and_datasets}

We utilize two public phonocardiogram (PCG) datasets. For in-domain training and evaluation, we use the \textbf{PhysioNet 2016 Challenge Dataset}~\cite{Liu_2016},~\cite{Goldberger2000}. The official split of PhysioNet is known to contain recordings from patients who are also present in the training set, creating a risk of data leakage. To ensure a scientifically valid evaluation, we identified and removed all such instances from our training data. For out-of-domain generalization testing, we use the \textbf{PASCAL Classifying Heart Sounds Dataset}~\cite{Bentley2011}]. We use the small, more challenging \textbf{Set B}, whose recordings were crowdsourced via the iStethoscope application, resulting in a significant domain shift from the cleaner, clinical-grade PhysioNet data. The train-test split was done based on the Patient IDs.

All models were trained for up to 200 epochs using the AdamW optimizer (learning rate $3 \times 10^{-4}$, weight decay $1 \times 10^{-4}$). Data imbalance was addressed through manual class weighting. We employed early stopping with a patience of 30 epochs on validation accuracy. We compare our proposed \texttt{AudioFuse} (concatenation fusion) against two single-modality baselines (a Spectrogram ViT and a Waveform 1D-CNN) whose architectures are identical to the branches of our fusion model. We also provide an ablation study on alternative Gated (FiLM) and Cross-Attention fusion, and a Spectrogram-Scalogram fusion model. To provide a comprehensive class-wise and overall evaluation, we report multiple metrics: Accuracy, F1-Score (abnormal), ROC-AUC, and the Matthews Correlation Coefficient (MCC).

\subsection{Performance on PhysioNet 2016}
\label{ssec:physionet_results}

The primary performance of all models was evaluated on the curated PhysioNet 2016 validation set. Table~\ref{tab:physionet_results_sota} summarizes these results. Our proposed \texttt{AudioFuse} model with concatenation fusion emerges as the top-performing architecture across almost all key metrics.

The single-modality baselines demonstrate competent but limited performance. The Raw Audio Baseline (ROC-AUC 0.8223) slightly outperforms the Spectrogram Baseline (0.8066), underscoring the importance of the temporal information preserved in the 1D waveform. The Spectrogram-Scalogram fusion performs poorly, indicating overfitting and that its input modalities are largely redundant.

\begin{table*}[t]
\centering
\vspace{-6 pt}
\caption{Model Performance Comparison on PhysioNet 2016. Results are reported as Mean $\pm$ Standard Deviation over multiple independent runs. Asterisks denote a statistically significant improvement over both baselines for that metric (*p $<$ 0.01).}
\label{tab:physionet_results_sota}
\small
\begin{tabular}{lccccc}
\toprule
\textbf{Model} & \textbf{\# Params} & \textbf{Accuracy} & \textbf{F1-Score} & \textbf{ROC-AUC} & \textbf{MCC} \\ 
\midrule
Spectrogram Baseline (ViT) & 1.83M & 0.7193 $\pm$ 0.0071 & 0.7383 $\pm$ 0.0197 & 0.8066 $\pm$ 0.0141 & 0.4444 $\pm$ 0.0211 \\
Raw Audio Baseline (1D-CNN) & 675K & 0.7376 $\pm$ 0.0094 & 0.7057 $\pm$ 0.0260 & 0.8223 $\pm$ 0.0313 & 0.4884 $\pm$ 0.0085 \\ \addlinespace
Spectrogram-Scalogram Fusion & 1.96M & 0.6611 $\pm$ 0.0215 & 0.6107 $\pm$ 0.0280 & 0.7478 $\pm$ 0.0251 & 0.3349 $\pm$ 0.0312 \\
AudioFuse (Cross-Attention) & 2.69M & 0.7276 $\pm$ 0.0152 & 0.7485 $\pm$ 0.0189 & 0.8226 $\pm$ 0.0161 & 0.4608 $\pm$ 0.0240 \\
AudioFuse (Gated FiLM) & 2.57M & 0.7575 $\pm$ 0.0110* & 0.7622 $\pm$ 0.0152* & 0.8518 $\pm$ 0.0141* & 0.5152 $\pm$ 0.0186* \\ 
\textbf{AudioFuse (Concatenation)} & 2.56M & \textbf{0.7741 $\pm$ 0.0094*} & \textbf{0.7664 $\pm$ 0.0005*} & 0.8608 $\pm$ 0.0127* & \textbf{0.5508 $\pm$ 0.0225*} \\
\midrule
MobileNetV3~\cite{Howard2019MobileNetV3} & 5.4M & 0.6677 $\pm$ 0.0165 & 0.6627 $\pm$ 0.0218 & 0.7509 $\pm$ 0.0204 & 0.3454 $\pm$ 0.0239 \\
EfficientNetB0~\cite{Tan2019EfficientNet} & 5.3M & 0.6467 $\pm$ 0.0041 & 0.6349 $\pm$ 0.0093 & 0.7472 $\pm$ 0.0177 & 0.4792 $\pm$ 0.0113 \\
ResNet18~\cite{He2016ResNet} & 11.7M & 0.7375 $\pm$ 0.0071 & 0.7365 $\pm$ 0.0080 & 0.8336 $\pm$ 0.0068 & 0.4792 $\pm$ 0.0113 \\
DenseNet169~\cite{Huang2017DenseNet} & 14.3M & 0.7464 $\pm$ 0.0284 & 0.7442 $\pm$ 0.0287 & \textbf{0.8640 $\pm$ 0.0181} & 0.5023 $\pm$ 0.0578 \\
InceptionV3~\cite{Szegedy2016InceptionV3} & 23.8M & 0.7621 $\pm$ 0.0181 & 0.7572 $\pm$ 0.0212 & 0.8581 $\pm$ 0.0089 & 0.5356 $\pm$ 0.0005 \\
\bottomrule
\end{tabular}
\end{table*}

While the more complex Gated FiLM and Cross-Attention fusion strategies show improvements over the baselines, they do not surpass the simple yet effective late-fusion concatenation. This suggests that for this task, a robust concatenation of specialized features is more stable and easier to optimize than more intricate fusion mechanisms. \texttt{AudioFuse} achieves an ROC-AUC of $0.8608 \pm 0.0127$, a significant improvement over the best-performing baselines.

Most importantly, despite being trained from scratch and having only 2.56M parameters, \texttt{AudioFuse} outperforms all the SOTA architectures (suitable versions for smaller datasets considered) trained on spectrograms in overall Accuracy, F1-Score, and Matthews Correlation Coefficient (MCC). Only DenseNet169 marginally beats AudioFuse in terms of ROC-AUC, but our model remains highly competitive with nearly 6 times fewer parameters.

Standard architectures, regardless of their depth or complexity, are fundamentally limited to the information present in a single view of the data, the spectrogram. \texttt{AudioFuse} circumvents this limitation. Its ViT branch is free to specialize in learning the global spectral patterns of the spectrogram, while the 1D-CNN branch independently captures the precise, high-resolution temporal dynamics from the raw waveform. The model is fed a richer, more disentangled set of features from the very same audio signal. It can simply learn more with less complexity. These results show that for rhythmic biomedical audio, our dual-representation fusion architecture can achieve better performance than a deeper, larger, single-input model even with a lightweight design.

\subsection{Domain Generalization Performance on PASCAL}
\label{ssec:pascal_results}

To test for the ability to generalize to unseen data from different recording environments, we evaluated our model on the PASCAL dataset using the weights trained solely on PhysioNet and fine-tuning the final dense layers. The results, shown in Table~\ref{tab:pascal_results}, highlight the robustness of \texttt{AudioFuse}.

The Spectrogram Baseline's performance collapses entirely, with an ROC-AUC of 0.4873. This suggests it overfitted to the specific acoustic characteristics of the PhysioNet data. The Raw Audio Baseline however, maintains a reasonable ROC-AUC of 0.6782, demonstrating that temporal features are less susceptible to this domain shift.

\begin{table}[th]
\centering
\vspace{-6 pt}
\caption{Model Performance on the PASCAL Dataset}
\label{tab:pascal_results}
\small 
\begin{tabular}{lcccc}
\toprule
\textbf{Model} & \textbf{Acc.} & \textbf{F1-Score} & \textbf{ROC-AUC} & \textbf{MCC} \\ 
\midrule
ViT & 0.5795 & 0.3273 & 0.4873 & 0.0579 \\
1D CNN & 0.6818 & 0.5484 & 0.6782 & 0.3152 \\
\textbf{AudioFuse} & \textbf{0.7386} & \textbf{0.6667} & \textbf{0.7181} & \textbf{0.4519} \\
\bottomrule
\end{tabular}
\end{table}

Once again, \texttt{AudioFuse} with Concatenation fusion achieves the best performance, with an ROC-AUC of 0.7181. This result is a dramatic improvement over both baselines. It demonstrates that by learning from two complementary representations, our model develops a more holistic and resilient understanding of the underlying cardiac sounds, making it far more generalizable and better suited for real-world applications where data variability is expected.

\section{Conclusion}
\label{sec:conclusion}

In this work, we demonstrated that a lightweight, hybrid architecture, \texttt{AudioFuse}, can achieve state-of-the-art competitive performance and superior generalization on PCG classification by simultaneously learning from complementary spectral (spectrogram) and temporal (waveform) representations, underlining the importance of utilizing multiple representations of signals. These results open up several promising avenues for future research. This approach can be directly extended to other audio domains where both tonal and transient features are critical. Furthermore, future work could explore scaling up these specialized branch architectures to leverage large-scale audio datasets, potentially rivaling the performance of heavily pre-trained models while retaining the benefits of a multi-representation learning approach.

\section{Code Availability}

The complete source codes for our baseline and AudioFuse models are publicly available on GitHub at: \url{https://github.com/Saiful185/AudioFuse}.

\vfill\pagebreak

\bibliographystyle{IEEEbib}
\bibliography{refs}

\end{document}